\documentclass[twocolumn,showpacs,preprintnumbers,amsmath,amssymb]{revtex4}
\usepackage{epsfig}
\usepackage{amsmath}

\begin{document}

\title{Trajectory Networks and Their Topological Changes Induced 
by Geographical Infiltration}

\author{Luciano da Fontoura Costa}
\affiliation{Instituto de  F\'{\i}sica de S\~ao Carlos, University of
S\~ao Paulo, P.O. Box 369, S\~ao Carlos, S\~ao Paulo, 13560-970
Brazil \\
and \\
Konrad Lorenz Institute for Evolution and Cognition Research,
Adolf Lorenz Gasse 2, A-3422 Altenberg, Austria}
\author{Cees van Leeuwen}
\affiliation{Laboratory of Perceptual Dynamics,
Riken, 2-1 Hirosava, Wako-shi, Saitama 351-0198 Japan \\}

\date{14th March 2008}

\begin{abstract}
We investigate the effect of progressive geographical infiltration on
the topology of trajectory networks.  Trajectory networks, a type of
knitted network, are obtained by establishing paths between
geographically distributed nodes while following an associated vector
field.  These systems offer tools for modeling adaptive growth,
development, and pathology of biological, transportation, or
communication networks.  For instance, the nodes could correspond to
neurons or axonal branching points along the cortical surface and the
vector field could correspond to the gradient of neurotrophic factors,
or the nodes could represent towns while the vector fields would be
given by economical and/or geographical gradients.  The geographical
infiltrations correspond to the addition of new local connections
between nearby existing nodes.  As such, these infiltrations could be
related to several real-world processes such as contaminations,
diseases, attacks, parasites, etc.  Combined with a mechanism for
elimination of nodes and connections, infiltration can model growth,
development and adaptive plasticity in neuronal networks. The
progressive geographical infiltration effect is expressed in terms of
the degree, clustering coefficient, size of the largest component and
the lengths of the existing chains measured along the infiltrations.
We show that the maximum infiltration distance plays a critical role
in the intensity of the induced topological changes.  For large enough
values of this parameter, the chains intrinsic to the trajectory
networks undergo a collapse which is unrelated to the percolation of
the network also implied by the infiltrations. (Copyright Luciano da
F. Costa, 2008)
\end{abstract}

\maketitle

\vspace{0.5cm}
\emph{`No one remembers what need or command or desire drove Zenobia's
founders to give their city this form, ..., which has perhaps grown
through successive superimpositions from the first, now undecipherable
plan.'
(I. Calvino, Inivisible Cities)}

\section{Introduction} 

Graphs and complex networks can be classified into two major
categories: \emph{geographical} and \emph{non-geographical} ones.
Whereas in the latter type of networks, nodes do not have specific
positions, in the former, each node has a well-defined spatial
position, expressible by respective coordinates.  Several real-world
networks are geographical in nature, including power distribution
(e.g.~\cite{Albert_power:2004}), tourism (e.g.~\cite{Baggio:2007}),
transportation (e.g.~\cite{Barthelemy:2006}), biological networks
(e.g. bone structure~\cite{Costa_bone:2006}, gene expression
expression~\cite{Costa_gene:2006, Costa_bioinf:2005}, and developing
neuronal networks~\cite{Bao:2008,vanPelt:2004}). They all share the
property that, to various extents, spatial proximity between nodes
plays a role in shaping the connectivity structure. Often in these
networks, spatially close nodes have a larger probability of being
connected. Sometimes the role of spatial position is more
intricate. For instance, in neuronal network development, axonal path
finding is directed by the cooperation of multiple factors.  These
include mechanical ones, such as the presence of a fissure, the
expression gradient of molecules as positive, permissive, or negative
guidance factors and the cell adhesion molecules involved in
fasciculation~\cite{Bao:2008}. In addition we need to
consider~\cite{vanPelt:2004} neurothropic factors that regulate
neuronal survival, differentiation, and
signaling~\cite{Huang:2001,Sofroniew:2001,Squire:2003}, the gradients
of neurotransmission~\cite{Semyanov:2005}, and the interactions
amongst these factors~\cite{Gong:2004,Kwok:2007}. To take these into
account, dynamical vector representations need to be associated with
network nodes, vertices, or geographical locations.  We wish to
incorporate these requirements into geographical networks.

A variety of geographical networks have been proposed in the
literature (e.g.~\cite{Kaiser_geo:2004, Costa_Sznajd:2005,
Gastner:2006, Warren:2002}).  A new family of networks, namely the
\emph{knitted networks}, was proposed recently~\cite{Costa_path:2007, 
Costa_comp:2007} to include all networks defined and composed by
paths, i.e. sequences of edges without repetition of nodes.

\begin{figure*}[htb]
  \vspace{0.3cm} 
  \begin{center}
  \includegraphics[width=0.4\linewidth]{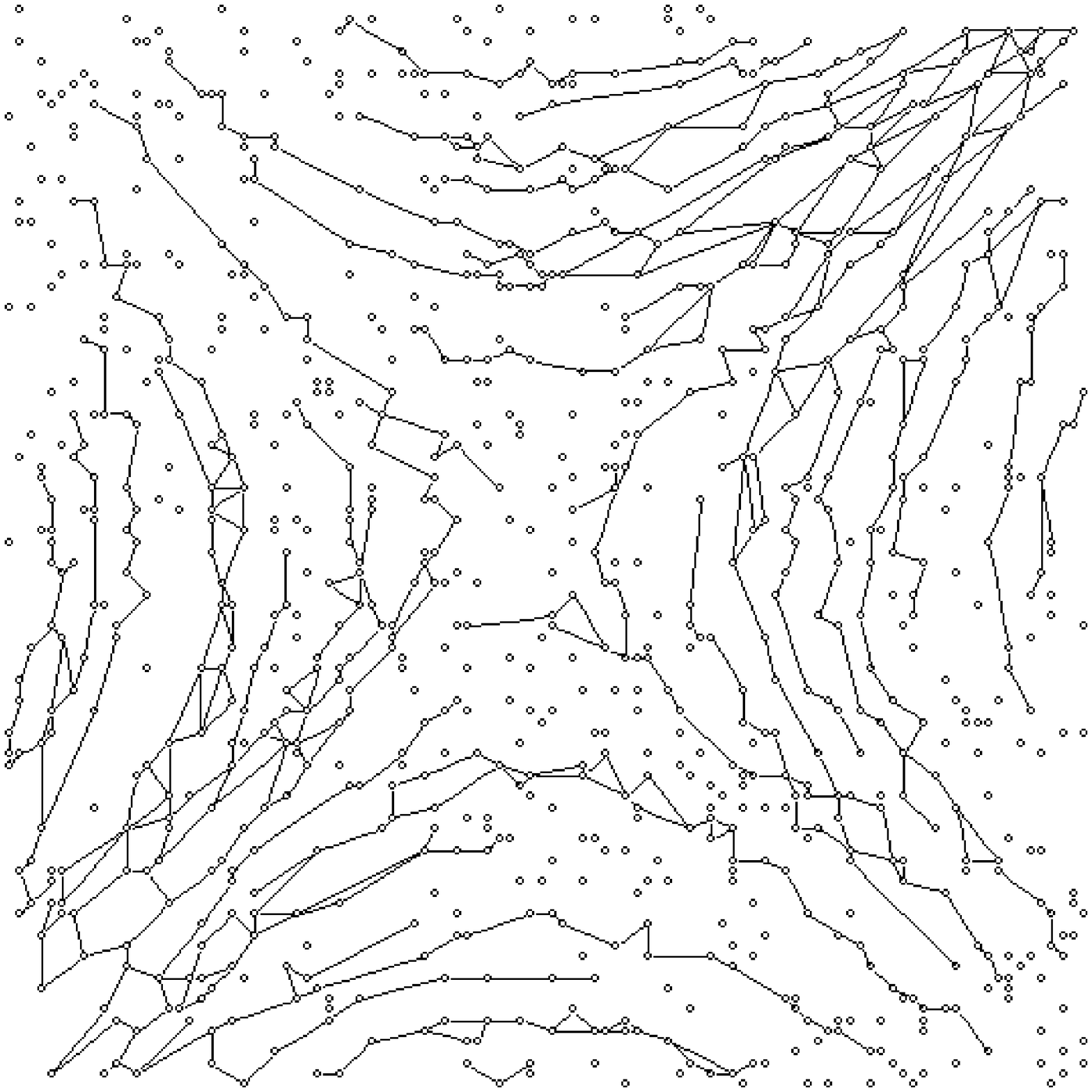}  
  \includegraphics[width=0.4\linewidth]{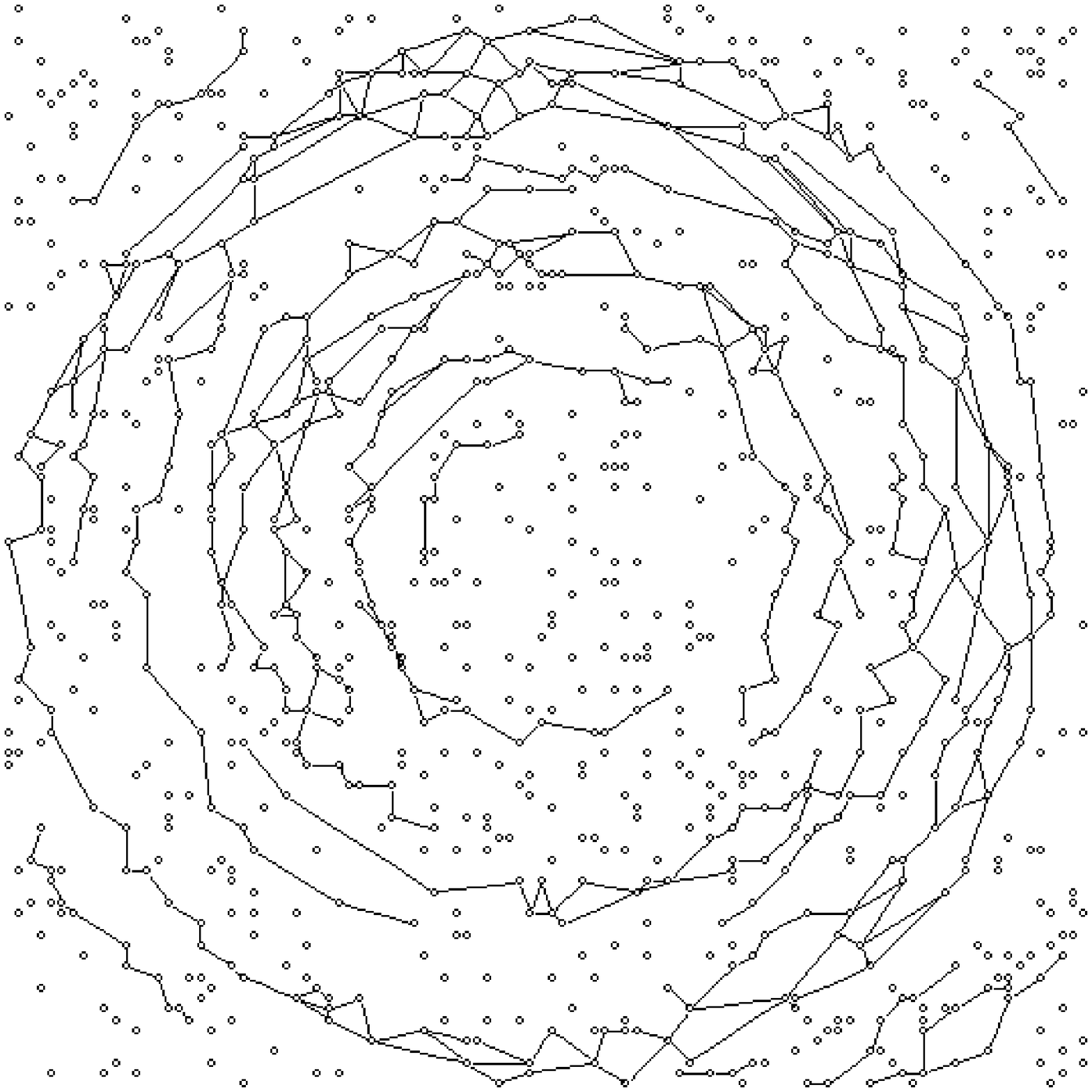} \\  
  (a)  \hspace{7cm}  (b) \\
  \caption{Trajectory networks obtained for the fields 
         $\vec{\phi}(x,y) = (y,x)$ (a) and $\vec{\phi}(x,y) = (y,-x)$ (b).
  }~\label{fig:ex} 
  \end{center}
\end{figure*}

In this article, we expand the family of knitted networks by
incorporating structures generated by trajectories defining paths
following a given vector field.  More specifically, a set of nodes is
distributed within a given domain (a 2D space in this article, but the
extension to higher dimensions is immediate); one node is chosen as
origin, and the respective trajectory (line of force) is obtained
while the nodes which are closer than a given maximum distance to the
current point of the trajectory are sequentially incorporated into the
path.  This procedure is repeated several times, yielding a network
with connections aligned to the vector field.  In other words, the
paths correspond to approximations of the solutions of the dynamical
system represented by the vector field.  Figure~\ref{fig:ex}
illustrates two trajectory networks obtained from the vector fields
$\vec{\phi}(x,y) = (y,x)$ and $\vec{\phi}(x,y) = (y,-x)$ (b).

Trajectory networks represent a natural putative model for several
real-world structures and phenomen, for instance neural growth and
development, including axonal navigation~\cite{Bao:2008}, the
establishment of neuronal connections under the influence of
neurotrophic fields (e.g.~\cite{Huang:2001, Sofroniew:2001,
Squire:2003}), neurotransmitter diffusion~\cite{Semyanov:2005} and
their relation with adaptive plasticity~\cite{Kwok:2007}, the growth
of transportation systems under geographical and economical influences
(e.g. `every path leads to Rome'), the growth of trees and roots under
influence of trophic factors~\cite{Gregory:2006}, the development of
channel-based systems such as bone structure and the vascular system,
amongst many other important systems.

The focus of attention in the current work is to investigate how the
topology of trajectory networks, a geographical type of knitted
network, is affected as a consequence of progressive
\emph{geographical infiltration}.  By geographical infiltration (hence
infiltration for short), we mean any process which interconnects pairs
of nodes.  Infiltration affects several real-world systems, e.g. the
appearance of cracks along channels, the establishment of new local
routes between towns and cities, contaminations between vessels of
fibers, gallery building by parasites, intentional attacks, internal
spreading of diseases, to cite just a few cases.  In the current work,
the infiltration process is simulated by selecting nodes at random and
connecting this node to all other nodes which are closer than a
maximum distance $D_p$.  Therefore, the adopted infiltration
corresponds to the progressive incorporation of \emph{tufts} of local
connectivity.

Here, we investigat the effects of progressive infiltration on the
topology of trajectory networks by quantifying the degree, clustering
coefficient, size of the largest component, as well as the number and
length of the chains present in the network.  A recent study
highlighted chains as an important category of network
motifs~\cite{Costa_chains:2008}.  Real-world networks often contain
several chains, in ways specific to their structure and function.
Thus, these networks are possibly the first theoretical model to
naturally incorporate these motifs.  These motifs are a consequence of
the linking of spatially distributed nodes along the trajectories
defined by the given vector fields.  The common trait in real-world
network structure that these models represent particularly well is the
presence of independent paths, with relatively few collaterals.
Therefore, it becomes particularly important to characterize the
structure of trajectory networks before and after infiltration by
considering the number and length of the existing chains.
Interestingly, the effect of infiltrations can be either bad or good,
depending on each specific system.  For instance, the incorporation of
additional local routes is in principle beneficial for transportation
and communication systems.  On the other hand, the addition of local
connections in biological networks (e.g. bone or neuronal networks)
may have catastrophic consequences.  Observe that in the latter
situation the main purpose of the chains/fibers is actually to provide
mutual isolation.  In both cases, the quantification of the effects of
the infiltration over the topology of the respective networks can
provide valuable information to be interpreted from the perspective of
each problem.

This article starts by presenting the basic concepts --- including the
generation of trajectory networks and the geographical infiltrations
--- and follows by describing the experiments and discussing the
respectively obtained results.

\section{Basic Concepts}

A \emph{complex network} is a graph exhibiting a particularly
intricate structure.  The connectivity of a undirected, unweighted
network can be completely represented in terms of the respective
\emph{adjacency matrix} $K$, such that each interconnection between 
two nodes $i$ and $j$ implies $K(i,j)=K(j,i)=1$, with
$K(i,j)=K(j,i)=0$ being otherwise imposed.  The \emph{immediate
neighbors} of a node $i$ are those nodes which receive an edge from
$i$.  The \emph{degree} of a node $i$ is equal to the number of its
immediate neighbors.  Two nodes are said to be \emph{adjacent} if they
share an edge; two edges are adjacent if they share one node.  A
sequence of adjacent edges is a \emph{walk}.  A \emph{path} is a walk
which never repeats a node or edge.  The length of a walk (or path) is
equal to the respective number of involved edges.  The
\emph{clustering coefficient} of nodes $i$ is calculated by dividing
the number of interconnections between its immediate neighbors and the
maximum possible number of connections which could be established
between those neighbors.

A \emph{connected component} of a network is a subgraph such that each
of its nodes can be reached from any of its other
nodes~\footnote{Often a connected component is understood to be
maximal, in the sense of incorporating all mutually reachable nodes in
that component.}.  A \emph{chain} is a subgraph of a network such as
that each of its nodes has degree 1 or 2 and not additional nodes of
degree 1 or 2 are connected to it~\cite{Costa_chains:2008}.  The
\emph{length} of a chain is given by its number of edges.  Two
measurements which can be used to characterize the chains in a given
network include the number of such chains and average and standard
deviation of their respective lengths.  Chains are naturally related
to paths along the network.

\section{Trajectory Networks}

A family of networks, namely the \emph{knitted complex networks}, was
introduced recently~\cite{Costa_path:2007, Costa_comp:2007}
incorporating all networks organized around the concept of
\emph{paths}.  Two main types of knitted networks were initially
identified: \emph{path-transformed} and
\emph{path-regular}.  The former subcategory of knitted complex
network is obtained by performing the start-path
transformation~\cite{Costa_path:2007} on a given network (star and
path connectivities can be understood as duals, e.g. through the
line-graph transformation).  Therefore, networks with power-law
distribution of path lengths can be obtained by star-path transforming
Barab\'asi-Albert networks~\cite{Albert_Barab:2002}.  The second type
of knitted complex networks, namely the path-regular networks, is
particularly simple and involves starting with a set of $N$ isolated
nodes and performing several paths encompassing all nodes.
Path-regular networks have been found to exhibit marked similar
properties between different configurations or nodes in the same
configuration (e.g.~\cite{Costa_comp:2007, Costa_longest:2007}).  An
even more regular version of the path-regular network, with all nodes
exhibiting identical degrees, was later reported
in~\cite{Costa_equiv:2008, Costa_conc:2008}.

Geographical networks are characterized by the fact that each of their
nodes has a well-defined spatial position.  Geographical networks
represent an important category of complex networks because several
real-world structures are inherently embedded into 2D or 3D spaces,
and their connectivities are strongly affected by proximity and
spatial adjacency.  Given a set of spatially distributed nodes
embedded in a continuous space to which a vector field is associated,
it is possible to obtain geographical networks whose connections are a
consequence not only of the proximity between nodes, but also of the
orientations implied by the respectively associated vector field.
Several real-world can be thought as involving a geographical
distribution of nodes and associated vector fields.  For instance, the
neurons along the cortical surface can be represented as a set of
geographically distributed nodes, while their connections are
established to a great extent as a consequence of neurotrophic fields
(e.g. electrical or chemical gradients).  Systems of streets, roads
and highways can also be understood as involving a set of spatially
distributed nodes (the intersections between routes), with the
interconnections being established in terms of the spatial proximity
between nodes as well as geographical and economical fields (e.g. the
trend to connect to a big city, to avoid a geographical obstacle or to
follow level-sets of height).  Several other natural and human-made
complex systems can be modeled by trajectory networks.  Trajectory
networks are related to gradient networks (e.g.~\cite{Toroczkai:2004,
Toro_Nature:2004, Costa_deriv:2005}), field
interactions~\cite{Costa_gene:2006, Costa_bioinf:2005,
Costa_saliency:2006}, as well as dynamical systems
(e.g.~\cite{Thurner:2005, Borges:2007}).  In the present work, we
understand trajectory networks as a particular case of knitted
networks.

The trajectory networks considered in the present article are obtained
as follows.  First, a two-dimensional workspace of size $L \times L$
is defined, and a vector field $\vec{\phi}(x,y)$ is associated to it.
For simplicity's sake we assume that $-L/2 \leq x,y \leq L/2$.  All
networks considered henceforth in this work are obtained for the
vector field $\vec{\phi}(x,y) = (y,x)$.  $N$ points are distributed
along this space with uniform probability.  A total of $N_p$
trajectories are then performed while obtaining each network.  A
starting point is randomly selected, and the respective line of force
(always parallel to the vector field) is calculated by using the Euler
leapfrog numerical method (e.g.~\cite{Mathews:1987}).  At each current
time, if a new node is found at a distance not exceeding $D_p$, that
node is connected to the previous node, and so on.  As it is clear
from the example of trajectory network shown in Figure~\ref{fig:ex},
the combination of proximity and orientation constraints while
performing the connections yield networks incorporating several
chains, which closely follow the vector field orientation.  Different
degrees of interconnectivity between and along the chains can be
obtained by varying the total number of points and the parameter
$D_p$.  Observe that the number of chains is is reduced for larger
values of $D_p/N$.  Once all trajectories are performed, the isolated
points can be removed (as adopted henceforth) or not (allowing further
connections).

\section{Geographical Infiltrations}

Given a geographical network, several types of perturbations of its
structure can arise as a specific consequence of its geographical
nature, in the sense that nodes which are spatially closer may
interfere one another.  For instance, in a neuronal system, unwanted
connections may appear between nearby neurons as a consequence of
diseases.  In transportation systems, it is only too natural to
incorporate new local connections to the network.  Several other types
of geographical interferences are possible, including those arising as
a consequence of contaminations, attacks, infiltrations, amongst many
other. 

In this work we incorporate progressive infiltrations to a given
network geographical network by selecting one of its nodes and
connecting to it all other nodes which are not further than a maximum
distance $D_i$.

\section{Results and Discussion}

\begin{figure*}[htb]
  \vspace{0.3cm} 
  \begin{center}
  \includegraphics[width=0.4\linewidth]{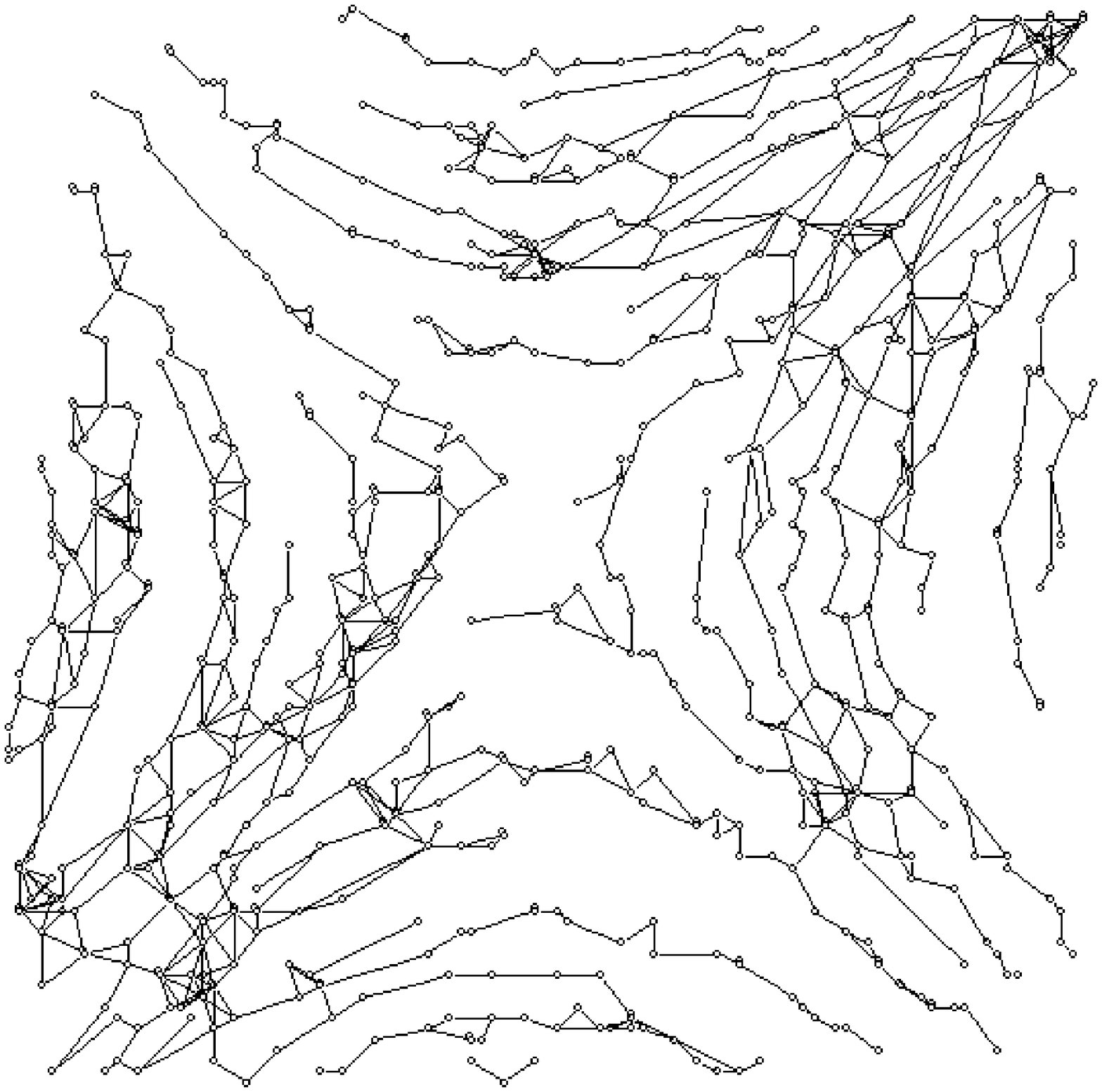}   
  \includegraphics[width=0.4\linewidth]{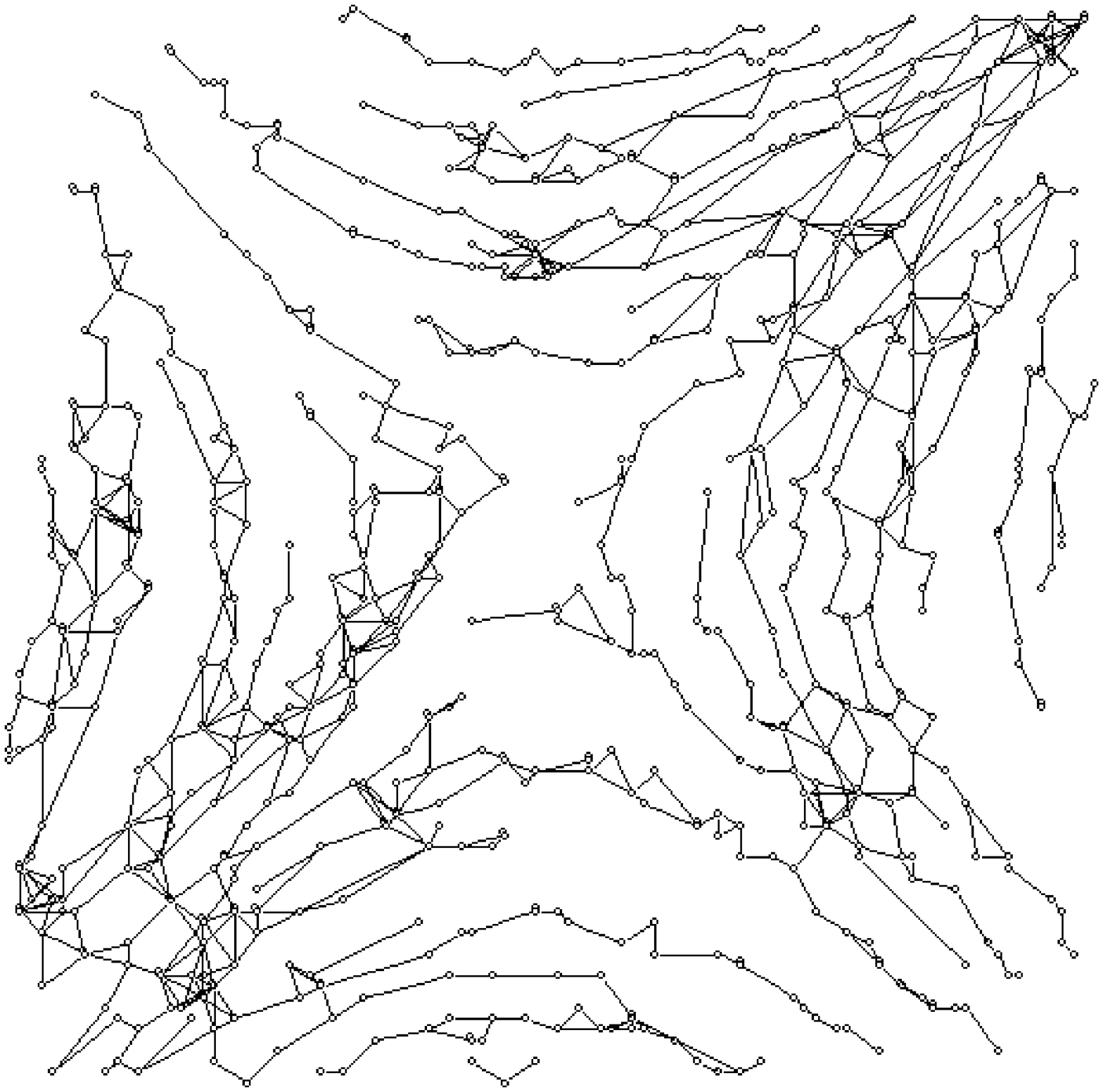}   \\
  (a) \hspace{7cm} (b) \\
  \includegraphics[width=0.4\linewidth]{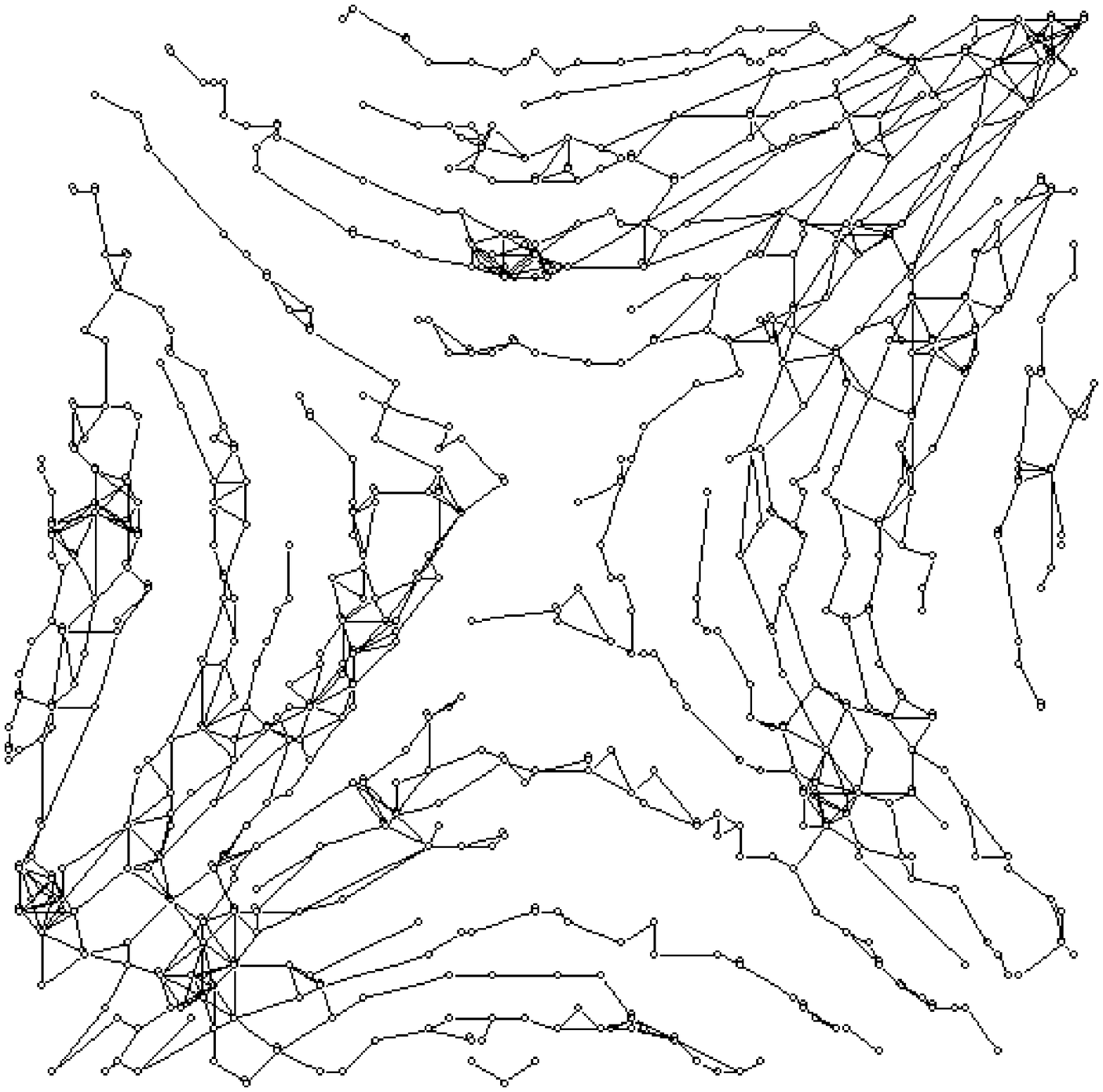}   
  \includegraphics[width=0.4\linewidth]{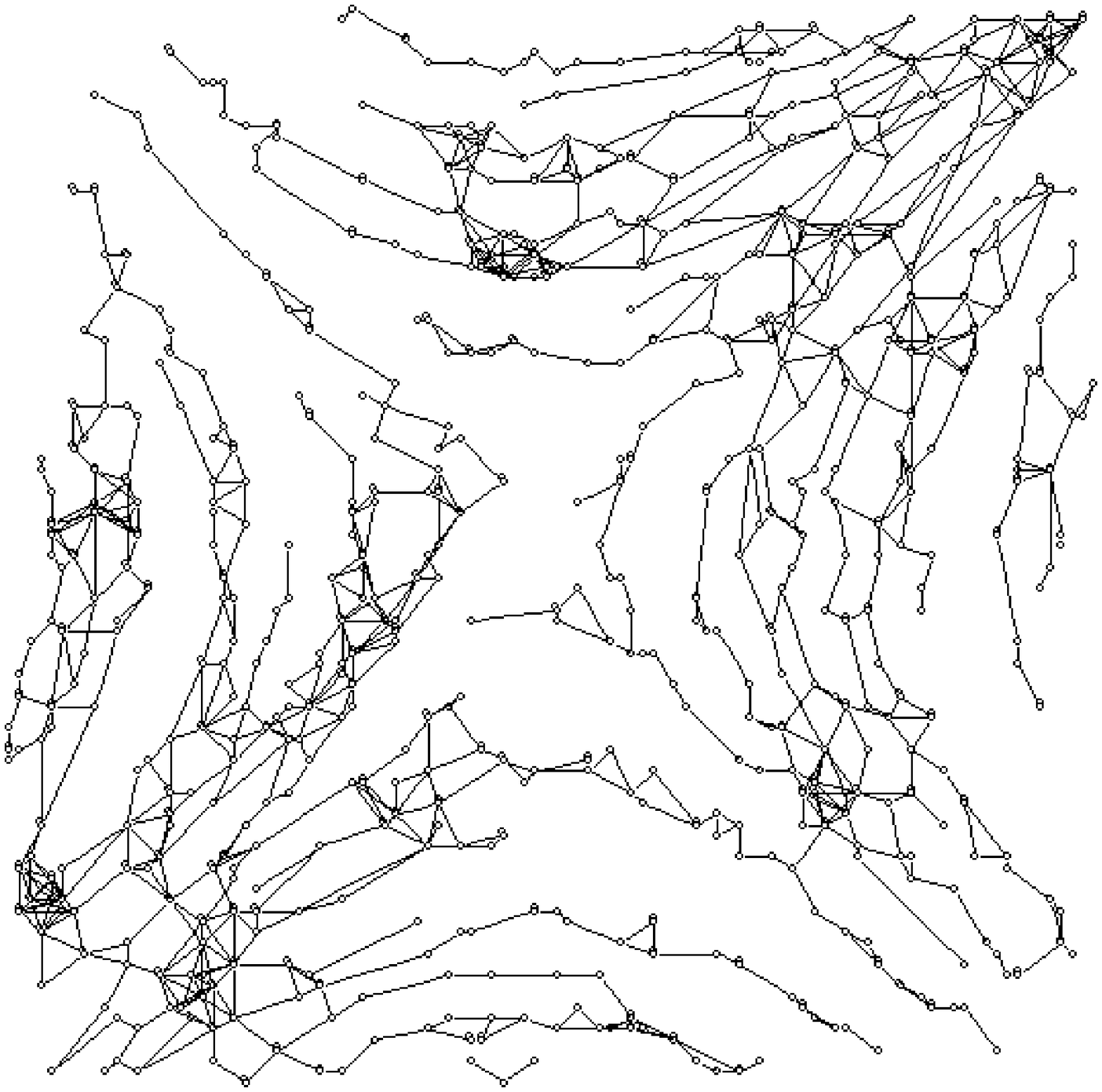}   \\
  (c) \hspace{7cm} (d) \\
  \caption{The network in Fig.~\ref{fig:ex} after 100 (a),
              200 (b), 300 (c) and 400 (d) infiltrations
              with $D_i = 5$. 
  }~\label{fig:inf_5} 
  \end{center}
\end{figure*}

\begin{figure*}[htb]
  \vspace{0.3cm} 
  \begin{center}
  \includegraphics[width=0.4\linewidth]{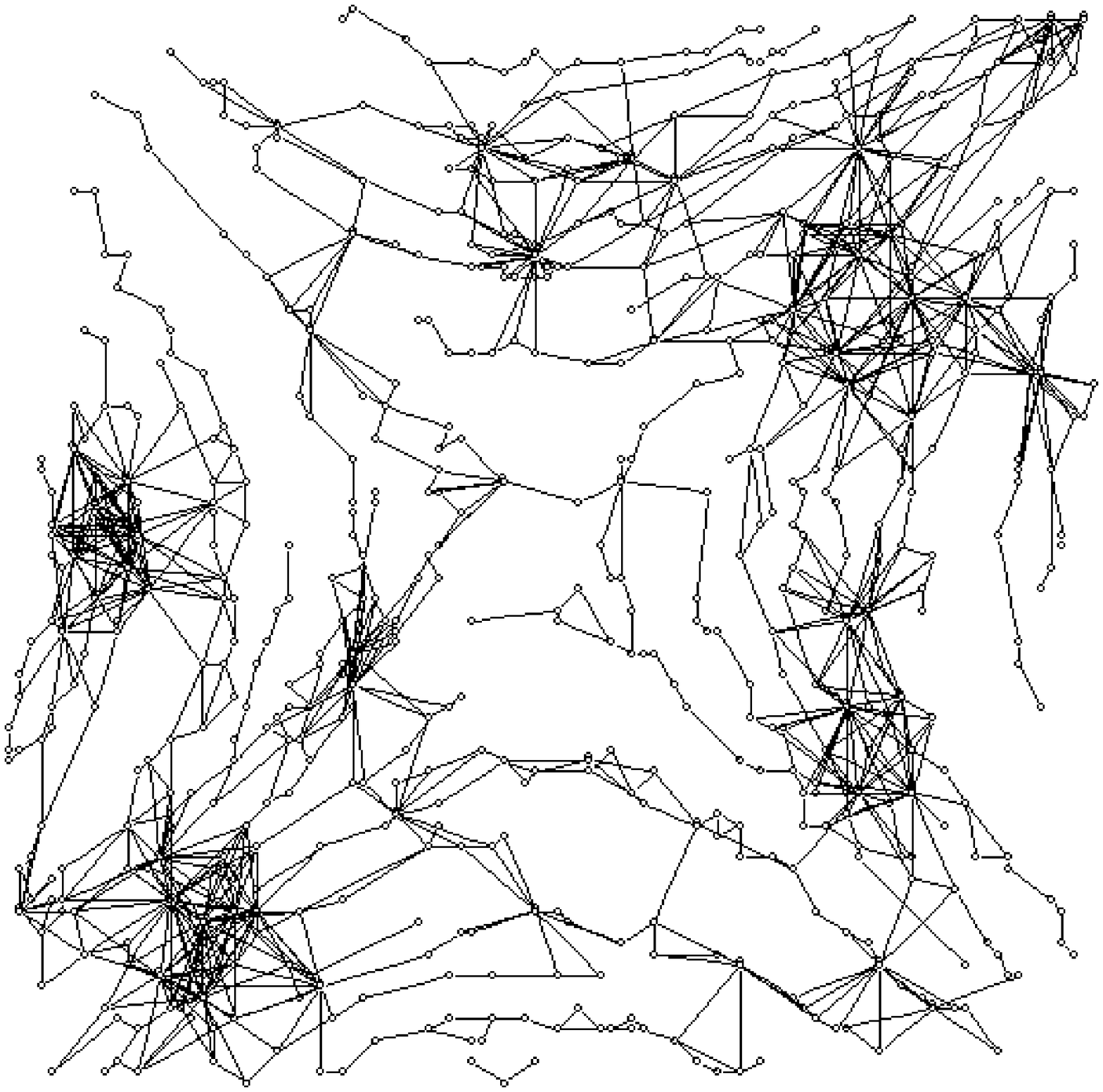} 
  \includegraphics[width=0.4\linewidth]{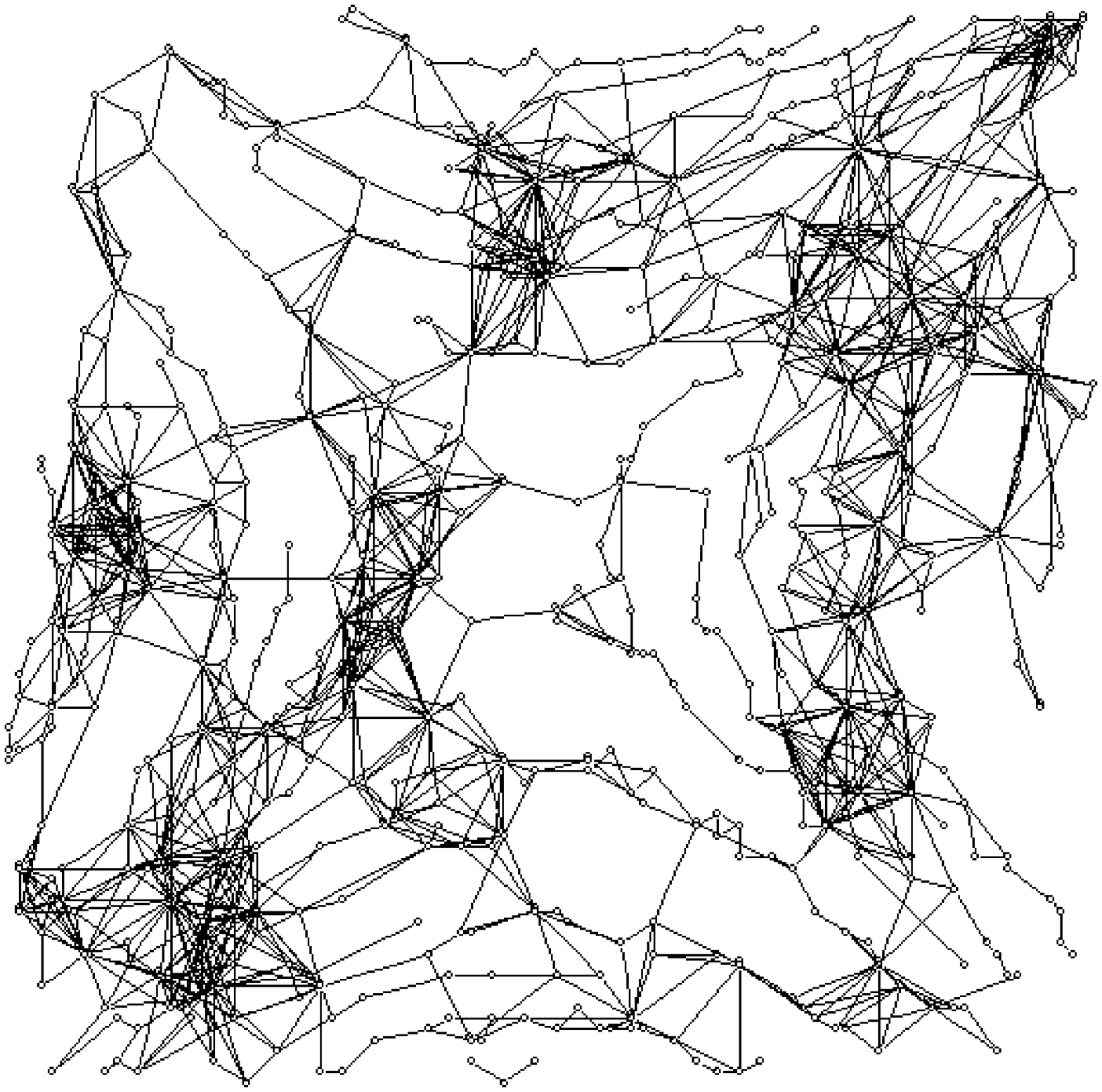}  \\
  (a) \hspace{7cm} (b) \\
  \includegraphics[width=0.4\linewidth]{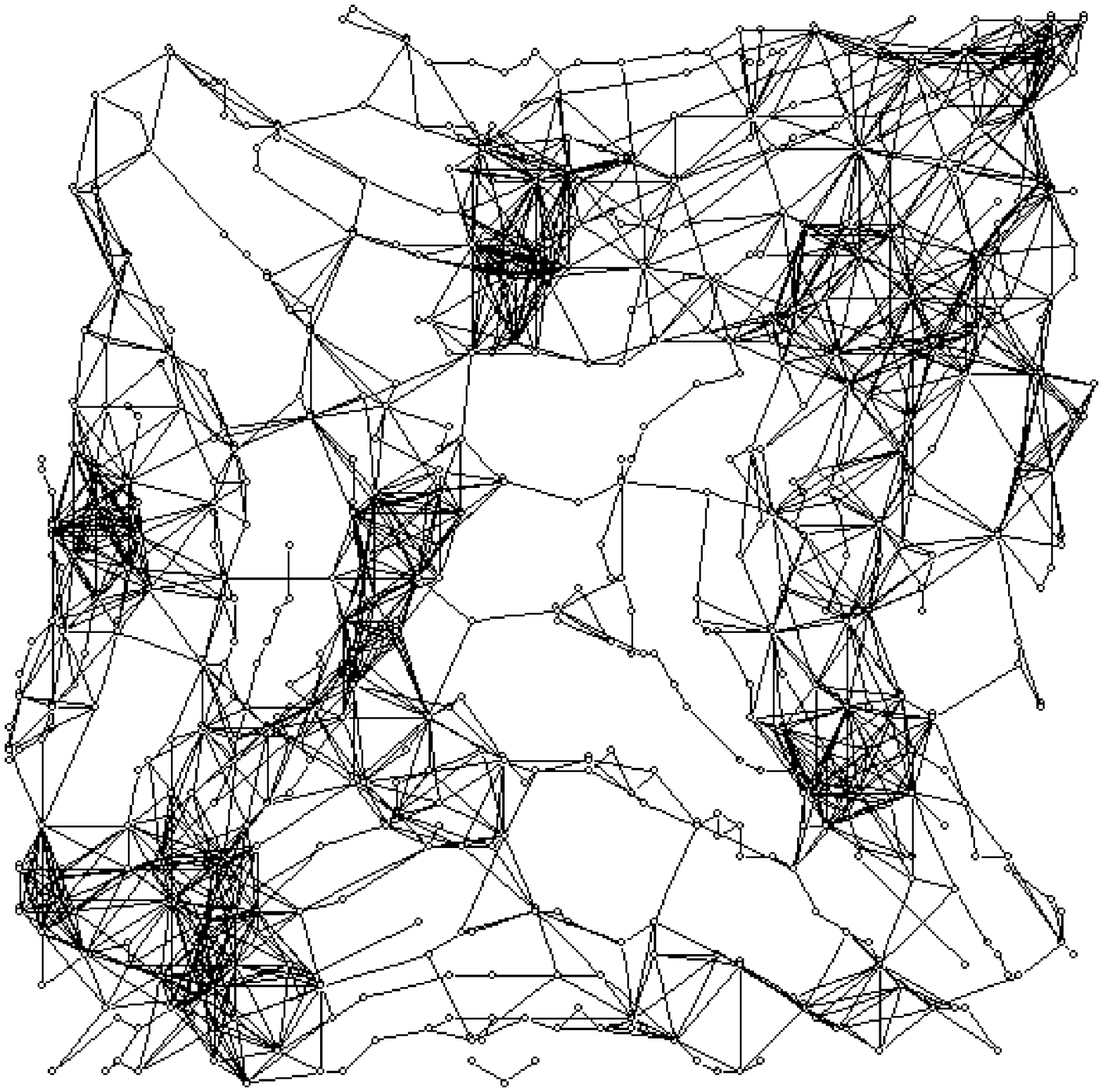} 
  \includegraphics[width=0.4\linewidth]{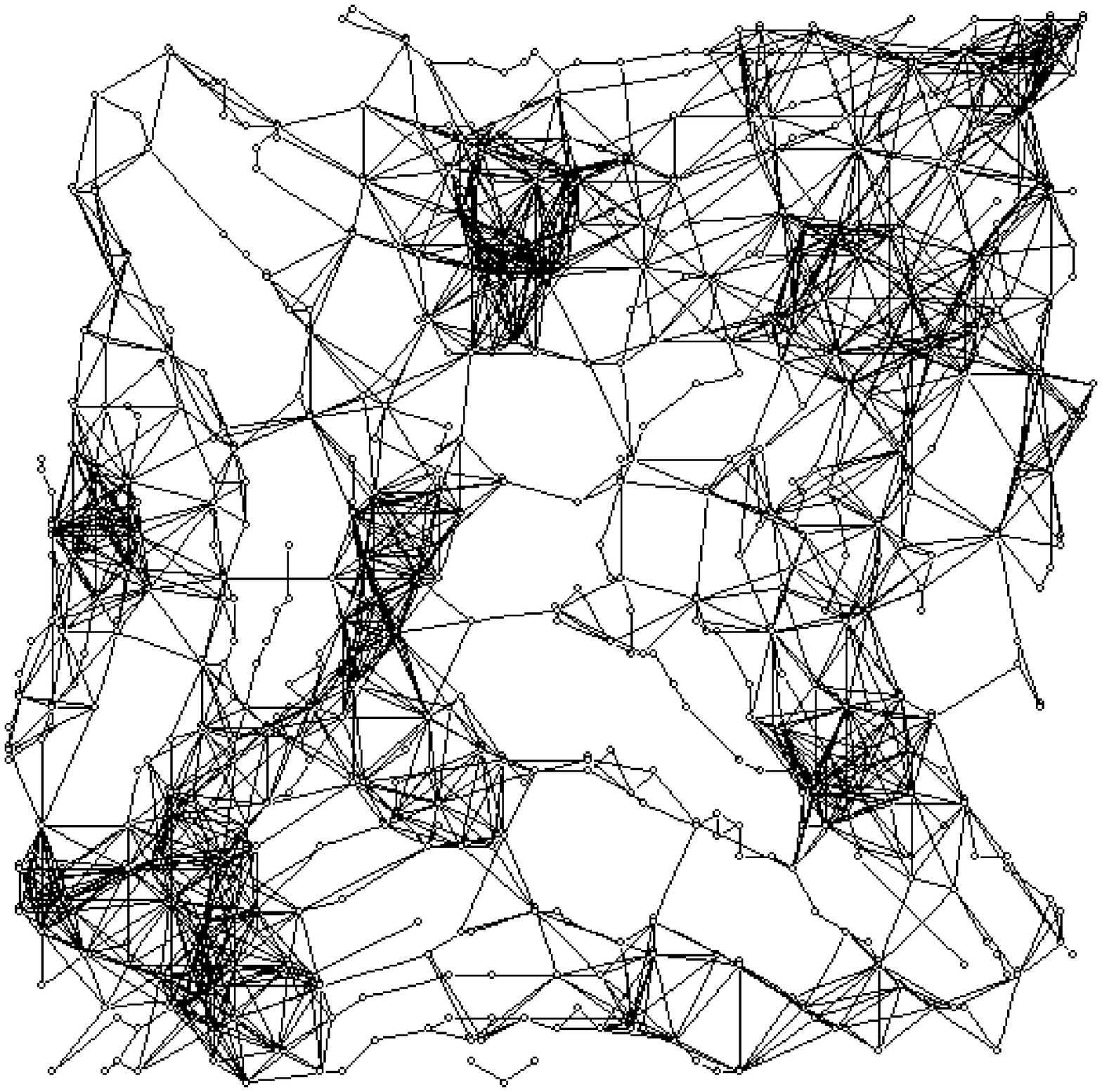} \\
  (c) \hspace{7cm} (d) \\
  \caption{The network in Fig.~\ref{fig:ex} after 100 (a),
              200 (b), 300 (c) and 400 (d) infiltrations
              with $D_i = 10$. 
  }~\label{fig:inf_10} 
  \end{center}
\end{figure*}

A set of 30 trajectory networks was obtained for the field
$\vec{\phi}(x,y) = (y,x)$.  A total of 1000 nodes was initially
distributed within a squre region of side $L=100$ centered at $(0,0)$,
and $N_p = 100$ trajectories were numerically calculated.  Starting
from a randomly chosen node, each node at a maximum distance $D_p=2$
from the current growing extremity of each trajectory was successively
connected.  An example of obtained trajectory network is shown in
Figure~\ref{fig:ex}.  Each of the 30 networks underwent progressive
infiltrations assuming $D_i=5$ and $D_i = 10$.  Figure~\ref{fig:inf_5}
shows four stages (100, 200, 300 and 400) along the successive
infiltrations for $D_i=5$.  Examples of the results of infiltrations
with $D_i = 10$ are depicted in figure~\ref{fig:inf_10}.

\begin{figure*}[htb]
  \vspace{0.3cm} 
  \begin{center}
  \includegraphics[width=0.9\linewidth]{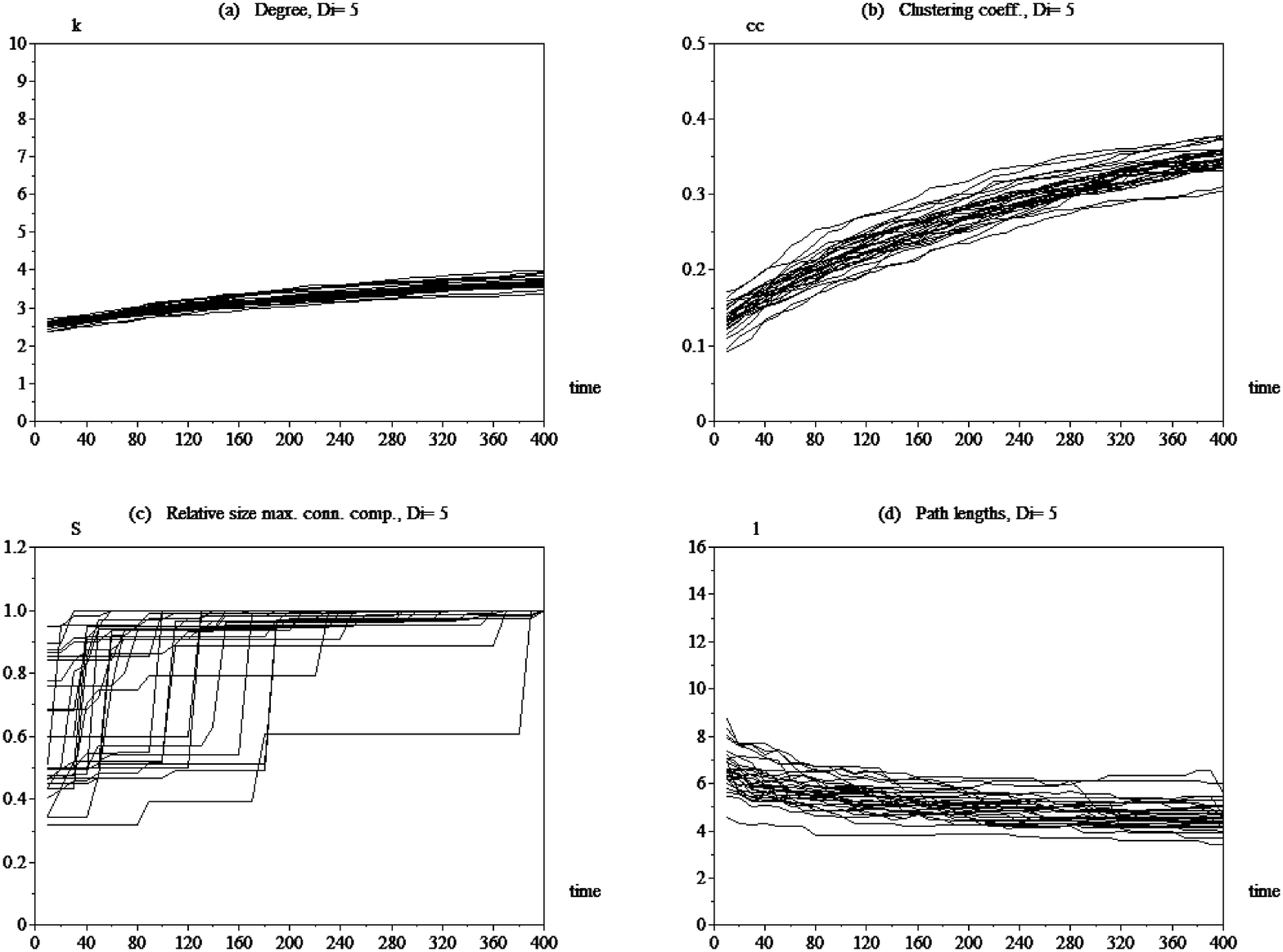} \\  
  \caption{Averages of degree, clustering coefficient, size of the
               largest connected component and chain lengths
               in terms of the number of infiltrations (identified as
               `time') with $D_i=5$ for each of the 30 networks 
               obtained for the vector field $\vec{\phi}(x,y) = (y,x)$.
  }~\label{fig:meas_all_5} 
  \end{center}
\end{figure*}

\begin{figure*}[htb]
  \vspace{0.3cm} 
  \begin{center}
  \includegraphics[width=0.9\linewidth]{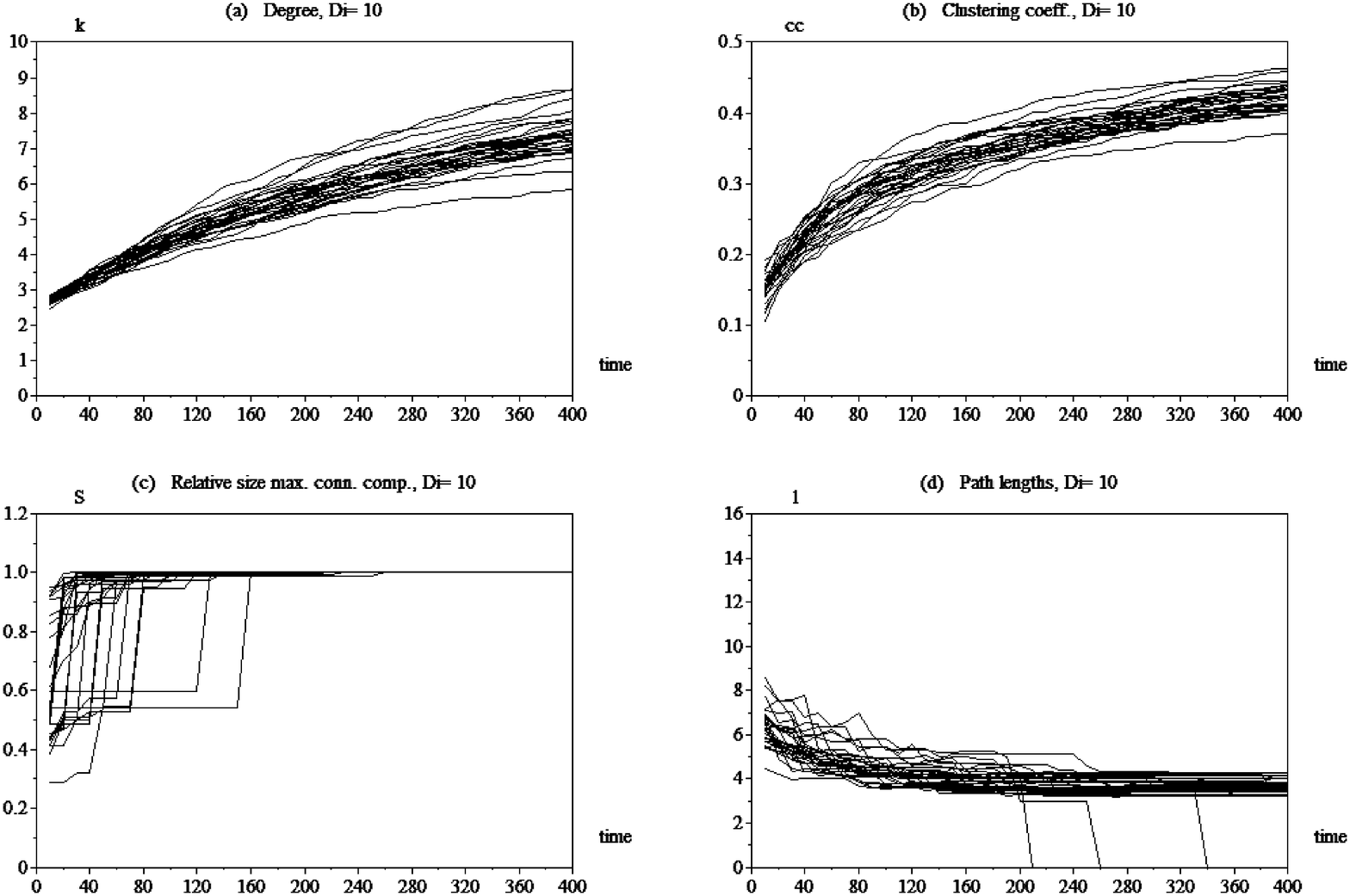} \\  
  \caption{Averages of degree, clustering coefficient, size of the
               largest connected component and chain lengths
               in terms of the number of infiltrations (identified as
               `time') with $D_i=10$ for each of the 30 networks 
               obtained for the vector field $\vec{\phi}(x,y) = (y,x)$.
  }~\label{fig:meas_all_10} 
  \end{center}
\end{figure*}

In order to characterize the alterations in the topology of the
trajectory networks as they underwent progressive infiltrations, a set
of measurements (e.g.~\cite{Costa_surv:2007}) was taken along the
process.  These measurements included the average and standard
deviation of the node degree, clustering coefficient, size of the
largest connected component, and chain lengths along successive
infiltration stages. Only chains longer than 3 edges were considered
in the respective measurements.  These chains were identified by
starting from each of the network nodes with degree 1 or 2 and
following along both sides (in case of degree 2) until the respective
extremities of the chains (nodes with degree 1 or larger than 2) were
found (each detected chain was removed from the network in order to
accelerate the processing of the remaining nodes).  The results
obtained for $D_i=5$ and $D_i=10$ are shown in Figure~\ref{fig:inf_5}
and~\ref{fig:inf_10}, respectively.  Figure~\ref{fig:meas_all_5}
and~\ref{fig:meas_all_10} show the above measurements for \emph{all} the
30 considered networks.

\begin{figure*}[htb]
  \vspace{0.3cm} 
  \begin{center}
  \includegraphics[width=0.9\linewidth]{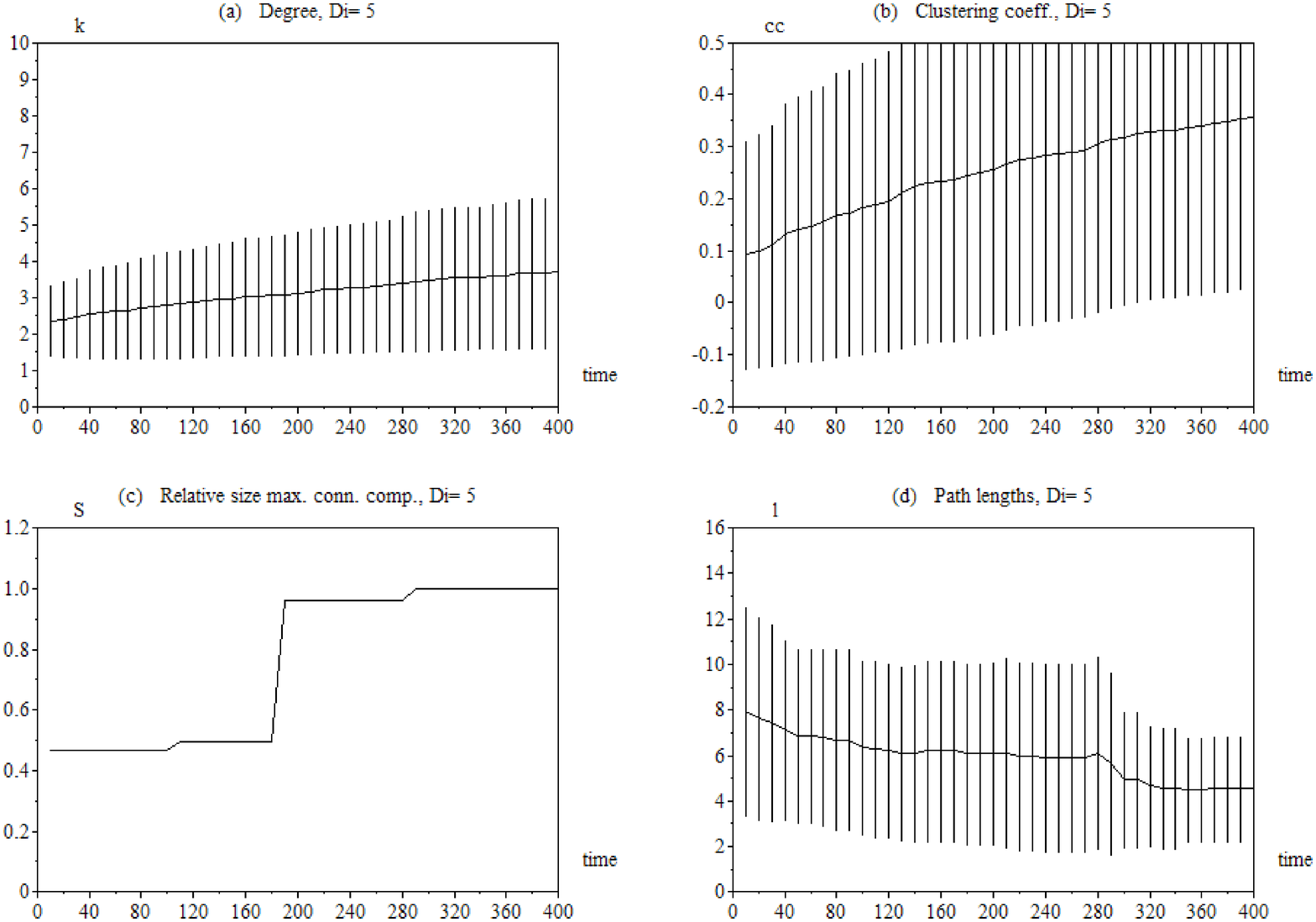} \\  
  \caption{Measurements of degree, clustering coefficient, size of the
               largest connected component and chain lengths
               in terms of the number of infiltrations (identified as
               `time') with $D_i=5$ for a network obtained for the 
               vector field $\vec{\phi}(x,y) = (y,x)$.
  }~\label{fig:meas_5} 
  \end{center}
\end{figure*}

\begin{figure*}[htb]
  \vspace{0.3cm} 
  \begin{center}
  \includegraphics[width=0.9\linewidth]{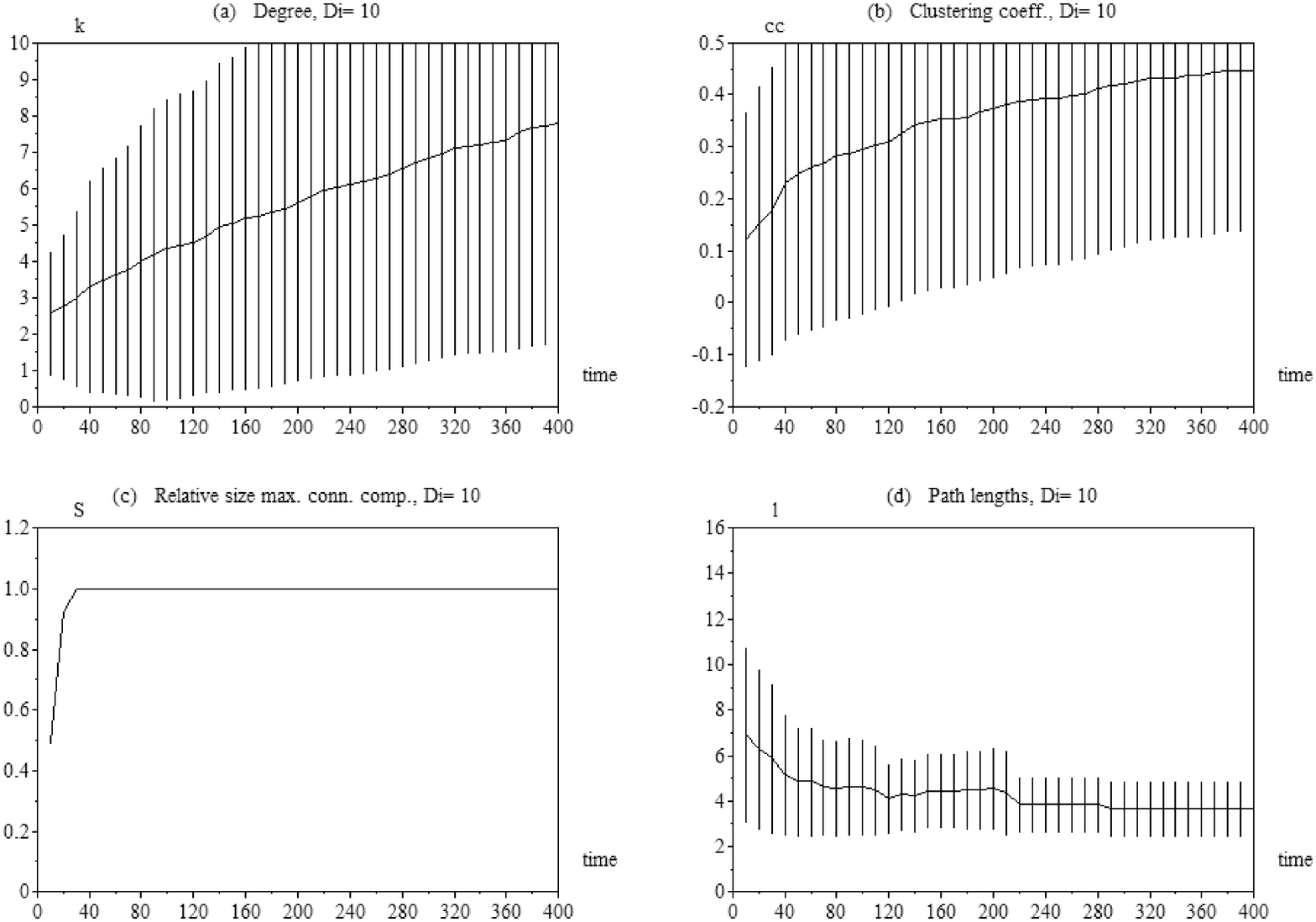} \\  
  \caption{Measurements of degree, clustering coefficient, size of the
               largest connected component and chain lengths
               in terms of the number of infiltrations (identified as
               `time') with $D_i=10$ for a network obtained for the 
               vector field $\vec{\phi}(x,y) = (y,x)$.
  }~\label{fig:meas_10} 
  \end{center}
\end{figure*}

\begin{figure}[htb]
  \vspace{0.3cm} 
  \begin{center}
  \includegraphics[width=0.9\linewidth]{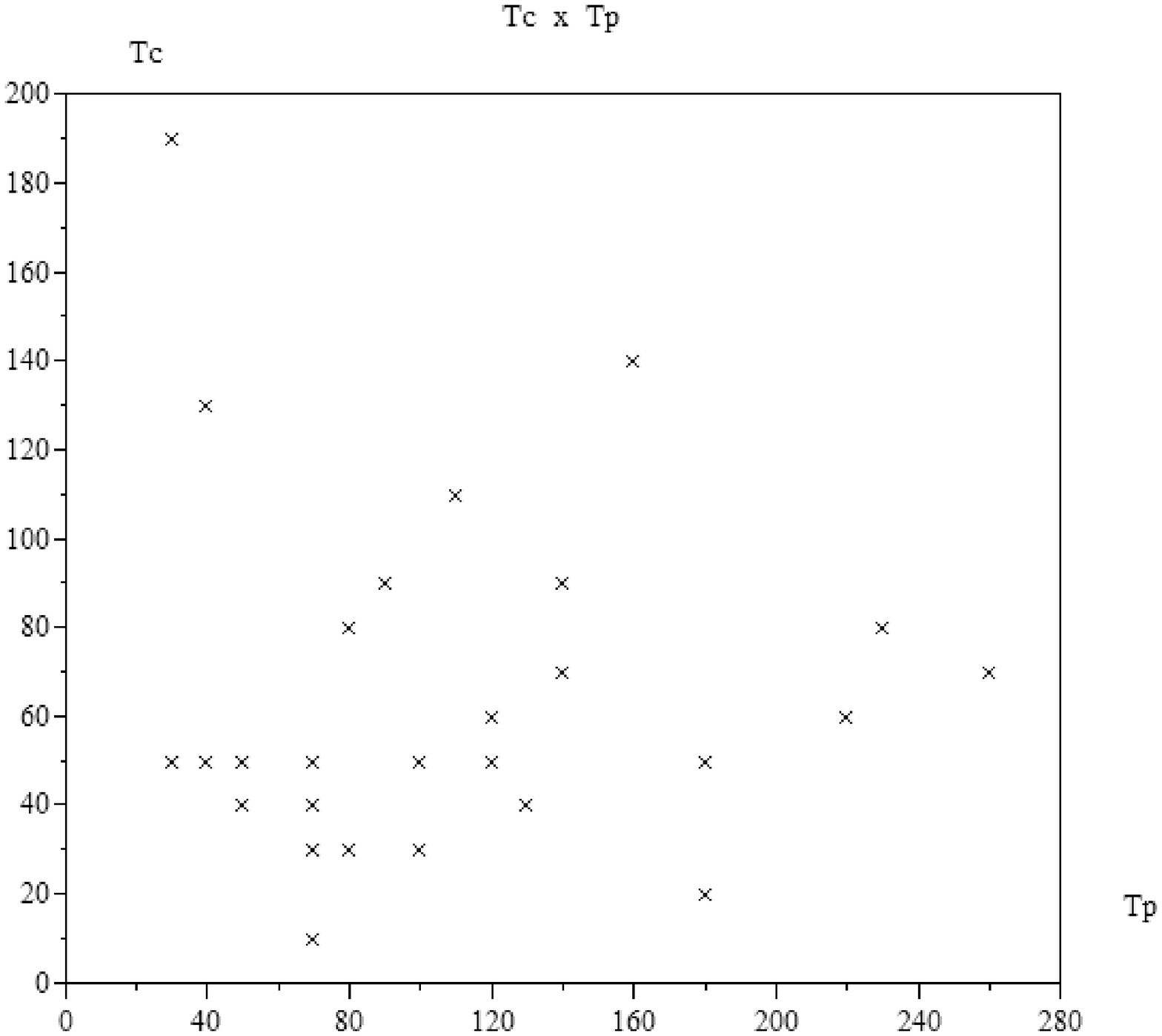} \\  
  \caption{Scatterplot of the percolation ($T_p$) and collapse 
             ($T_c$) critical times for $D_i=10$.
  }~\label{fig:scatt} 
  \end{center}
\end{figure}

It is clear from Figures~\ref{fig:meas_5} to~\ref{fig:meas_all_10}
that, as could be expected, the degree and clustering coefficient both
increased as a consequence of the addition of the infiltration tufts.
Both such increases are sublinear, with a steeper decrease in the rate
of clustering coefficient increase observed for $D_i=10$
(Fig.~\ref{fig:meas_all_10}).  The relative sizes of the maximum
connected components suffer an abrupt transition before the 160 first
infiltrations (most of the transitions take place before that value)
for both settings of $D_i$, but is more abrupt for $D_i=10$.  This
change is related to the percolation of the chains in the original
network.  Another relatively abrupt change is observed for the path
lengths, most of which stabilizing themselves at a value near 6 for
$D_i=5$ and 4 for $D_i=10$.  The interval from the start of the
infiltrations until the average length of the chains stabilizes (as
observed above) is called the \emph{period of collapse} of the chains.
Very few networks remained with large average chain lengths larger
after 200 infiltrations.  This confirms the fact, evident from
Figure~\ref{fig:scatt}, that the tuft infiltrations tend to quickly
eliminate most of the long chains in the trajectory networks (the
chain collapse).  For larger values of $D_i$, after the collapse of
the chains, the vector field influence on the network connectivity can
be hardly distinguished by visual inspection, such as in
Figures~\ref{fig:inf_10}(b-d).  It is important to keep in mind that
the fact that small values of $D_i$ tend to imply little effect over
the chain structure of the trajectory networks is ultimately related
to the number $N$ of initial nodes and the maximal distance $D_p$
considered for chaining the nodes during the construction of the
networks.

The two involved critical phenomena, namely the percolation of the
networks and the collapse of the chains, were investigated further in
order to search for possible relationship between their respective
onsets.  In order to do so, transition points along the successive
infiltrations were identified automatically.  These points,
respectively $T_p$ and $T_c$, correspond to the first occurrence of
the value 1 for the relative size of the largest connected component
and the first occurrence of the average chain length which is smaller
or equal than 5, respectively.  Figure~\ref{fig:scatt} shows the
respectively obtained distribution of $T_p$ and $T_c$ obtained for the
30 realizations of networks with $D_i=10$.  It is clear from this
figure that the two critical phenomena taking place in the considered
trajectory networks seem to be largely independent, in the sense that
no correlation has been observed between their critical values.
Interestingly, as shown in Figure~\ref{fig:scatt}, the collapse of the
chains can take place before the respective percolation.

\section{Concluding Remarks}

Geographical networks represent an important category of complex
networks because of their natural potential for modeling a large
number of real-world and human-made complex structures and systems.
At the same time, the category of complex networks built up by paths,
namely the \emph{knitted networks}, constitutes an important
superclass of complex structures because of their intrinsic
association with the concept of paths (as opposited to star
connectivity) and random walk dynamics (e.g.~\cite{Costa_path:2007,
Costa_comp:2007, Costa_longest:2007}).  In this work, trajectory
networks have been understood to belong to the supercategory of
knitted networks as a consequence of the fact that these structures
are the result of path generation processes.  Trajectory networks
constitute a special case, in which the paths tend to follow an
associated vector field. Our main interest in the present work,
however, consisted in investigating how the topology of trajectory
networks changed as a consequence of geographical infiltrations.
While several types of attacks and perturbations have been considered
and investigated in complex network research, relatively less
attention has been devoted to perturbations intrinsically related to
geographical constraints, especially the adjacency and proximity
between nodes.  Yet, several important real-world and human-made
systems are prone to this type of perturbations, ranging from the
onset of unwanted neuronal connections to the incorporation of new
local routes to transportation systems.

The main contributions reported in this article are listed and
reviewed in the following:

\emph{Trajectory networks as a special case of knitted complex networks:} 
We have defined trajectory networks as a novel sub-class of knitted
networks.  This type of geographical knitted network corresponds to an
interesting case where the connectivity is the consequence of both the
proximity between nodes and the orientation of the underlying vector
field.

\emph{New type of perturbation of network structure:} We considered, 
possibly for the first time, perturbations (or `attacks') to
geographical networks which depend on the proximity between the
spatially distributed nodes.  We focused attention on `tuft'
infiltrations, where a node $i$ is randomly chosen and all other nodes
which are closer than a maximum distance $D_i$ are connected to node
$i$.  This type of topological change can be related to several
real-world effects such as unwanted neuronal tangles as a consequence
of diseases, establishment of local connections in transportation
networks, contaminations, and attacks.

\emph{Qualitative changes resulting from infiltrations:} The progressive 
infiltration of a trajectory network was investigated in a systematic
manner, considering 30 realizations of networks obtained for the same
configuration with respect to the vector field $\vec{\phi}(x,y) =
(y,x)$.  The changes in the networks topology was monitored by taking
several measurements including the degree, clustering coefficient,
size of the largest connected component, as well as the particularly
relevant lengths of the existing chains.  The latter measurements are
especially important because the trajectory networks are inherently
composed by chains.  While the degree and clustering coefficients
underwent relatively smooth increases, the size of the largest
component and average chain lengths were subjected to relatively
abrupt changes related to the percolation of the network (in the case
of the largest connected component) and to the collapse of the chain
structure (in the case of the average chain lengths).  The value of
$D_i$ was found to be have great influence on such topological changes
induced by the infiltrations, with values much larger than $D_p$
implying particularly intense changes, especially regarding the chain
structure.  After the collapse of the chains, the effect of the
original vector field on the network connectivity could hardly be
discerned.  Such findings are particularly important for a large
number of real-world structures underlain by trajectory networks and
geographical infiltrations.

\emph{Independence of percolation and collapse of chains:}  The 
progressive infiltration of trajectory networks involves two critical
phenomena: its percolation and the collapse of its chain structure.
Interestingly, no clear relationship between these phenomena has been
identified by considering the critical times $T_p$ and $T_c$.  This
implies that the collapse of the chains can not be predicted from the
percolation of the respective network, and vice-versa.  As a matter of
fact, it has also been observed that the collapse of the chains can
take place before the percolation of the respective network.

\emph{Modeling of Brain Development} The current study illustrates that, 
in order to avoid pathological network conditions, besides growth, a
mechanism for the selective elimination of connections is also
necessary. Such a mechanism can be observed at work in brain
development. The formation of neuronal networks involves the extensive
growth, but also elimination of neurons and
connections~\cite{Quartz:99}. Isolated nerve cells undergo apoptosis;
dendritic arbors are being built and retracted based on signaling
efficacy and electrical activity in the pre and postsynaptic
neurons. As a result, synapses undergo extensive rewiring after their
initial attachment~\cite{Zhang_Poo:2001}. These processes work
together to maintain a functional network architecture for effective
communication between brain cells~\cite{Kwok:2007}.

The several possibilities of future work include but are not limited
to the following:

\emph{Other types of vector fields:}  It would be interesting to 
investigate how the patterns of topological changes observed in this
work extends to trajectory networks obtained by considering other
vector fields, as well as other configurations of the involved
parameters.

\emph{Orthogonal infiltrations:} In this work we focused attention on  
tuft infiltrations.  It would be interesting to study the topological
changes of trajectory networks with respect of other types of
geographical perturbations, such as connecting points according to
proximity and orientations orthogonal to the vector field (possibly
also through trajectories).

\emph{Infiltration by increasing distances:}  While the infiltrations 
implemented in this article consisted in selecting nodes followed by
tuft interconnection, it would be particularly interesting to
investigate the topological alterations of trajectory networks while
all pairs of nodes are joined according to successive distances.  Such
a type of infiltration is guaranteed to completely eliminate the
chains after a critical interval.

\emph{Application to real-world networks:}  It would be interesting to 
quantify the alterations of real-world networks expressible by
trajectory networks, including transportation networks, power
distribution, communications, tourism and neuronal systems.

\emph{Application to Image and Shape Analysis:}  The analysis of images 
containing objects and shapes has remained a great challenge
(e.g.~\cite{Costa_book:2001,Costa_saliency:2006}).  It would be
particularly interesting to consider the application of the concepts
and methods reported in the current work to such problems. More
specifically, trajectories can be obtained in gray-level images by
considering their respective gradient fields.  So, by distributing
points through the image and interconnecting them while taking in to
account trajectories driven by the gradient fields, it is possible to
obtain respective network representations incorporating a great deal
of the intrinsic geometric features.  Shapes represented by their
contour can also be mapped into trajectory networks by considering
vector fields induced by their borders (e.g. electrical or distance
fields).  The topological properties of the respective measurements
are expected to provide valuable features for image and shape analysis
and classification.  Signatures obtained by considering the evolution
of several measurements of the so-obtained networks as the consequence
of geographical infiltration can provide additional features for
visual characterization and classification.

\begin{acknowledgments}
Luciano da F. Costa thanks CNPq (301303/06-1) and FAPESP (05/00587-5)
for sponsorship.
\end{acknowledgments}

\bibliography{ginfil}
\end{document}